\documentclass[preprint2]{aastex}

\shorttitle{Water Masers in NGC 1333 IRAS 4}
\shortauthors{Park \& Choi}

\newcommand{\HtO}{{\rm H$_2$O}}
\newcommand{\kms}{\mbox{km s$^{-1}$}}% km/s

\begin{document}

\title{Observations of Water Masers in the NGC 1333 IRAS 4 Region}
\author{\sc Geumsook Park\altaffilmark{1,2,3}
            and Minho Choi\altaffilmark{1}}
\altaffiltext{1}{International Center for Astrophysics,
                 Korea Astronomy and Space Science Institute,
                 Hwaam 61-1, Yuseong, Daejeon 305-348, South Korea.}
\altaffiltext{2}{Department of Astronomy and Space Science,
                 Chungnam National University, Daejeon 305-764, South Korea.}
\altaffiltext{3}{pgs@kasi.re.kr}
\setcounter{footnote}{3}

\begin{abstract}
The NGC 1333 IRAS 4 region was observed in the 22 GHz H$_2$O maser line
with an angular resolution of about 0.08 arcseconds.
Two groups of masers were detected,
one near IRAS 4A and the other near BI.
Among the eight maser spots detected near IRAS 4A,
six spots are located close to A2, within 100 AU,
and the maser velocities are
also near the systemic velocity of the cloud core.
These masers are probably related with the circumstellar disk.
Since there is no maser spot detected around A1,
the star forming process is relatively more active in A2 than in A1.
Four maser spots were detected near IRAS 4BI.
Since most of them are distributed along a straight line
in the direction of the outflow,
BI masers are most likely related with the jet or outflow.
The disk-outflow dichotomy of H$_2$O masers is discussed briefly.
No maser was detected near H$_2$O(C),
another maser source reported previously.
\end{abstract}


\section{INTRODUCTION}

NGC 1333 is a very active star forming region
at a distance of 320 pc from the Sun (Jennings et al. 1987;
de Zeeuw et al. 1999).
IRAS 4 is a multiple-star system, consisting of at least five young
stellar objects (YSOs) (Sandell et al. 1991; Rodr{\'\i}guez et al. 1999;
Looney et al. 2000).
IRAS 4A1 and A2 form a relatively close binary system (Lay et al. 1995;
Looney et al. 2000).
IRAS 4BI is located $\sim$30$''$ southeast of IRAS 4A.
They are Class 0 protostars with strong submillimeter emission
(Sandell et al. 1991).
The nature of the other objects, IRAS 4BII and IRAS 4C, are less clear.

IRAS 4A is a protobinary system with a separation of 1\farcs8 or 580 AU
(Looney et al. 2000; Reipurth et al. 2002).
A1 is brighter than A2 in the radio continuum emission
(Looney et al. 2000; Reipurth et al. 2002;
Girart et al. 2006; Choi et al. 2007).
However, the outflow driven by A2 is stronger and larger than that of A1
(Choi 2005; Choi et al. 2006).
The northeastern outflow of A2 has an intriguing bent feature
that was interpreted as a result of collision between the outflow and
a dense cloud core (Choi 2005).
Recently, circumstellar disks were discovered by NH$_3$ observations
(Choi et al. 2007).
BI was suspected to be a multiple system by Lay et al. (1995).
Later, however, imaging observations showed BI to be a single object.
(Looney et al. 2000; Reipurth et al. 2002).
The interpretation of Lay et al. (1995) could have been a confusion
caused by BII.
BI has a bipolar outflow in the north-south direction (Choi 2001).

Haschick et al. (1980) detected a water maser spot,
\HtO(C), using a single-dish telescope.
Since \HtO(C) is located near the bending point of the northeastern outflow
(see Figure 4 of Choi 2005),
they could be related.
Later, several maser spots were detected around A2 and BI
(Rodr{\'\i}guez et al. 2002; Furuya et al. 2003).
However, \HtO(C) was not detected by these interferometric observations.

To study the water masers in the two protostellar systems,
IRAS 4A and IRAS 4BI,
and to find the exact location of \HtO(C),
we observed the IRAS 4 region using the Very Large Array (VLA).
In \S~2 we describe our observations.
In \S~3 we report the results.
In \S~4 we discuss the implications of our results and
the relation between the water masers and the protostar system.
A summary is given in \S~5.

\section{OBSERVATIONS}

The NGC 1333 IRAS 4 region was observed
using the VLA of the National
Radio Astronomy Observatory\footnote{The NRAO is a facility of the National
Science Foundation operated under cooperative agreement by Associated
Universities. Inc.}
in the $6_{16} \rightarrow 5_{23}$ transition of \HtO\ (22.235077 MHz).
The observations were made in two tracks in the A-array configuration.
Twenty-two antennas were used on 2006 March 24, and
twenty-three antennas were used on 2006 April 24.
The spectral windows were set to have 128 channels
with a channel width of 0.024 MHz,
giving a velocity resolution of 0.33 \kms.
The phase and bandpass calibrator was the quasar 0336+323 (4C 32.14).
The flux calibration was done by observing the quasar 0713+438 (QSO B0710+439).
The flux density of 0713+438 was 0.47 Jy in March and 0.44 Jy in April,
which are the flux densities measured within a day of our observations
(VLA Calibrator Flux Density Database\footnote{See
http://aips2.nrao.edu/vla/calflux.html.}).
Comparison of the amplitude gave a flux density of 1.00 Jy in March
and 0.92 Jy in April for 0336+323,
which agrees with the value in the VLA Calibrator Flux Density
Database to within 9\% and 2\%, respectively.
To avoid the degradation of sensitivity owing to pointing errors,
pointing was referenced by observing the calibrators in the $X$-band (3.6 cm).
This referenced pointing was performed at the beginning of each track
and just before observing the flux calibrator.

The phase tracking center was
$\alpha_{2000}$ = 03$^{\rm h}$29$^{\rm m}$10\fs529
and $\delta_{2000}$ = 31\arcdeg13$'$31\farcs000.
Imaging was done with a uniform weighting,
and maps were made using a CLEAN algorithm.
The synthesized beams were
FWHM = 0\farcs090 $\times$ 0\farcs075 and P.A. = $-41\arcdeg$ in March,
and FWHM = 0\farcs094 $\times$ 0\farcs077 and P.A. = $-53\arcdeg$ in April.
The maps have rms noises of 2.4 and 2.5 mJy beam$^{-1}$, respectively.
The absolute positional accuracy of the VLA under normal conditions
is expected to be about 0\farcs1 in the A-array configuration.
The continuum emission was not detected.
Figure 1 shows the known compact sources in our field of view.

\section{RESULTS}

Eight maser spots were detected around IRAS 4A2,
and four spots were detected near IRAS 4BI.
All the A2 masers are redshifted with respect to the
systemic velocity of the cloud core ($V_{\rm LSR}$ = 6.7 \kms;
Blake et al. 1995; Choi 2001).
Out of the four maser spots near BI, two spots are redshifted and
the others are blueshifted with respect to the systemic velocity
of the cloud core ($V_{\rm LSR}$ = 6.6 \kms; Blake et al. 1995).
We examined the whole field of view, but
no maser source was detected around \HtO(C).
Table 1 lists the source parameters,
and Figures 2 and 3 show the distribution of the detected \HtO\ maser spots.
Spectra of the \HtO\ masers are shown in Figure 4.

\section{DISCUSSION}

Out of the twelve \HtO\ maser spots detected,
eight are associated with IRAS 4A2, and four with IRAS 4BI.
Previous observations with VLA also showed several \HtO\ maser spots.
BI masers were detected by both
Rodr{\'\i}guez et al. (2002) and Furuya et al. (2003), and
A2 masers were detected by Furuya et al. (2003) only.

\subsection{IRAS 4A2}

Six maser spots, PC 2--5, 7, and 8,
are located near the radio continuum source A2, within 100 AU.
The maser velocity ranges from 7.2 to 8.8 \kms,
and the maser lines are near the systemic velocity (6.7 \kms;
Blake et al. 1995; Choi 2001), within about 2 \kms.
The tight positional association with the continuum source
and the small velocity differences suggest
that these maser spots may be related
with the circumstellar disk, not the outflow.
These maser emission probably comes from shocked gas in the accretion disk.
The exact nature of the shock is not clear.
There are other examples of \HtO\ masers
that are probably emitted by the shocked gas in protostellar disks,
including IRAS 00338+6312, IC 1396N, and NGC 2071 IRS 1/3
(Fiebig et al. 1996; Slysh et al. 1999; Seth et al. 2002).

PC 3 and 5 in March are identical to
PC 7 and 8 in April, respectively, in both position and velocity (Table 1).
These two spots seem to have a relatively long lifetime,
at least a month.
Since the maser line detected by Furuya et al. (2003) have
different velocities, the lifetime is shorter than a few years.

PC 1 and 6 are relatively far from A2, at least 300 AU.
Since they are located in the direction of the bipolar outflow,
these maser spots are probably related with the jet or outflow.

No maser spot associated with A1 was detected,
which probably indicate that the A2 disk is more active than the A1 disk.
This interpretation is consistent with the facts
that A2 drives more powerful outflow than A1
and that the A2 disk is brighter in NH$_3$ lines than the A1 disk
(Choi 2005; Choi et al. 2007).

\subsection{IRAS 4BI}

The systemic velocity of the BI cloud core is $\sim$6.6 \kms\
(Blake et al. 1995),
and the maser velocity ranges from $-1.2$ to 19.9 \kms\
(Rodr{\'\i}guez et al. 2002; this work).
The maser spots lie along a 0\farcs7--long straight line
in the southeast-northwest direction, P.A. $\approx\ 29\arcdeg$ (Fig. 3).
The HCN outflow observed by Choi (2001) flows in a similar direction.
These facts suggest that the BI masers
are emitted from the shocked gas related with the outflow.
Figure 5 shows the position--velocity diagram,
which shows that the blueshifted masers
tend to be located in the southeast direction,
while the redshifted ones in the northwest.
This trend agrees with the HCN outflow.

\subsection{Dichotomy between Disk and Outflow}

Torrelles et al. (1997, 1998) claimed that 
\HtO\ masers often prefer to trace selectively either the disk
or the outflow in a specific region
and proposed that this dichotomy may be caused by
the differences in the evolutionary stage of the driving source.
They suggested that the \HtO\ maser traces the disk in younger sources
and the outflow in older sources. 

Our observations appear to show
that such a dichotomy may exist in the NGC 1333 IRAS 4 region. 
Most of the A2 masers trace the disk, while the BI masers trace the outflow.
However, the measured outflow timescales suggest
that A2 ($\sim$2000 yr; Choi et al. 2006)
is much older than BI ($\sim$120 yr; Choi 2001). 
Therefore, the evolutionary explanation of the dichotomy
by Torrelles et al. (1997, 1998) does not seem to hold.

Then what could cause the apparent dichotomy?
A possible explanation can simply be
the low occurrence rate of the maser phenomenon.
The \HtO\ line amplification requires
the existence of warm high-density shocked gas (Elitzur 1995),
and \HtO\ maser in low-mass YSOs is a rare phenomenon.
Previous surveys of \HtO\ masers toward low-mass YSOs showed
that the detection rate is typically 5--20 \% 
(Wilking \& Claussen 1987; Terebey et al. 1992;
Persi et al. 1994; Furuya et al. 2003; G{\'o}mez et al. 2006).
Since the detection probability
of either disk-maser or outflow-maser is small,
the probability of detecting both disk- and outflow-masers
around a single YSO would be even smaller.
For example, let us assume
that the detection probability of \HtO\ maser from a disk is 10 \%,
and that the probability is similar for an outflow.
If the two kinds of maser occur independently,
the probability of detecting both in a specific region
would be only $\sim$1 \%.
That is, in this example, out of all the regions associated with masers,
95 \% of the sample would show only one kind of masers,
either disk-maser or outflow-maser, not both.
Therefore, the dichotomy could be a result of low probability.

\subsection{H$_2$O(C)}

The maser source \HtO(C) reported by Haschick et al. (1980) was not detected
in any of the observations with VLA
(Rodr{\'\i}guez et al. 2002; Furuya et al. 2003; this work).
There are several possible reasons for the nondetection.
First, all the VLA observations might be unfortunately
made during inactive phases of \HtO(C).
Second, \HtO(C) might be a mis-identification of A2.
Haschick et al. (1980) observed using a single-dish telescope
with a large beam size.
The right ascension of \HtO(C) coincides with that of A2,
and the declination of \HtO(C) could be in error.

\acknowledgments

We thank J. Cho for helpful discussions and encouragement.
This work was partially supported by the LRG Program of KASI.
G. P. was partially supported by the BK 21 project of the Korean Government.

\begin{deluxetable}{p{19mm}llrrcccc}
\tabletypesize{\footnotesize}
\tablecaption{H$_2$O Maser Sources in the NGC 1333 IRAS 4 Region}
\tablewidth{0pt}
\tablehead{
 & \multicolumn{4}{c}{\sc Peak Position} & \colhead{} &
\colhead{} & \colhead{\sc Peak Flux\tablenotemark{d}}\\
\cline{2-5}
\colhead{\sc Source\tablenotemark{a}} & \colhead{$\alpha_{2000}$} &
\colhead{$\delta_{2000}$} & 
\colhead{$\Delta \alpha$\tablenotemark{b}} & 
\colhead{$\Delta \delta$\tablenotemark{b}} &
\colhead{$V_{0}$\tablenotemark{c}} &
\colhead{$\Delta V$\tablenotemark{c}} &
\colhead{(Jy beam$^{-1}$)}
}
\startdata
PC 1\dotfill &03 29 10.490 &31 13 32.683 &--0.497 \phn $\pm\ 0.004$\phn
& 1.683  \phn $\pm\ 0.004\phn$ & 8.7 &0.7 &\phn0.022 $\pm\ 0.002$ \\
PC 2\dotfill &03 29 10.404 &31 13 32.454 &--1.598 \phn $\pm\ 0.004$\phn
& 1.454 \phn $\pm\ 0.004$\phn& 8.8 &0.8 &\phn0.023 $\pm\ 0.002$ \\
PC 3\dotfill &03 29 10.4174&31 13 32.3733&--1.4317 $\pm\ 0.0001$
&1.3733 $\pm\ 0.0001$ & 8.7 &0.8 &\phn0.735 $\pm\ 0.002$ \\
PC 4\dotfill &03 29 10.401 &31 13 32.287 &--1.645 \phn $\pm\ 0.005$\phn
& 1.287  \phn $\pm\ 0.005$\phn& 8.8 &1.1 &\phn0.019 $\pm\ 0.002$ \\
PC 5\dotfill &03 29 10.422 &31 13 32.246 &--1.373 \phn $\pm\ 0.004$\phn
& 1.246  \phn $\pm\ 0.004$\phn& 7.3 &0.8 &\phn0.024 $\pm\ 0.002$ \\
PC 6\dotfill &03 29 10.313 &31 13 30.235 &--2.768 \phn $\pm\ 0.005$\phn
&--0.765  \phn $\pm\ 0.005$\phn& 8.8 &1.0 &\phn0.020 $\pm\ 0.002$ \\
PC 7\dotfill &03 29 10.417 &31 13 32.377 &--1.433 \phn $\pm\ 0.001$\phn
& 1.377  \phn $\pm\ 0.001$\phn& 8.8 &1.1 &\phn0.121 $\pm\ 0.003$ \\
PC 8\dotfill &03 29 10.422 &31 13 32.238 &--1.370 \phn $\pm\ 0.002$\phn
& 1.238  \phn $\pm\ 0.002$\phn& 7.2 &0.8 &\phn0.044 $\pm\ 0.003$ \\
PC 9\dotfill &03 29 12.007 &31 13 08.008 &18.955 \phn $\pm\ 0.004$\phn
&--22.992 \phn $\pm\ 0.004$\phn& 1.2&0.7 &\phn0.031 $\pm\ 0.003$ \\
PC 10\dotfill&03 29 12.0128 &31 13 08.9457 &19.0339 $\pm\ 0.0001$
&--23.0543 $\pm\ 0.0001$&--1.2&0.6&   40.973 $\pm\ 0.003$ \\
PC 11\dotfill&03 29 11.9951 &31 13 08.3301 &18.8073 $\pm\ 0.0002$
&--22.6699 $\pm\ 0.0002$& 19.9&1.0 &\phn0.588 $\pm\ 0.003$ \\
PC 12\dotfill&03 29 11.937 &31 13 06.724 &18.057 \phn $\pm\ 0.008$\phn
&--24.276 \phn $\pm\ 0.008$\phn& 19.8&1.1 &\phn0.024 $\pm\ 0.003$ \\
\enddata
\tablecomments{Units of right ascension are hours, minutes, and seconds,
               and units of declination are degrees, arcminutes,
               and arcseconds.
               All velocities are in \kms.}
\tablenotetext{a}{\HtO\ maser spots presented in this work
                  are labeled with a prefix PC.}
\tablenotetext{b}{Position offsets in arcseconds 
                  relative to the phase tracking center (see \S~2). 
                  Positional uncertainties were estimated
                  by considering the signal-to-noise ratios
                  and the pixel size.}
\tablenotetext{c}{The central velocity and the line width (FWHM)
                  estimated by a Gaussian fit to each spectrum.}
\tablenotetext{d}{Fluxes were corrected for the primary beam response.}
\end{deluxetable}

\clearpage

\begin{figure}
\includegraphics[width=75mm]{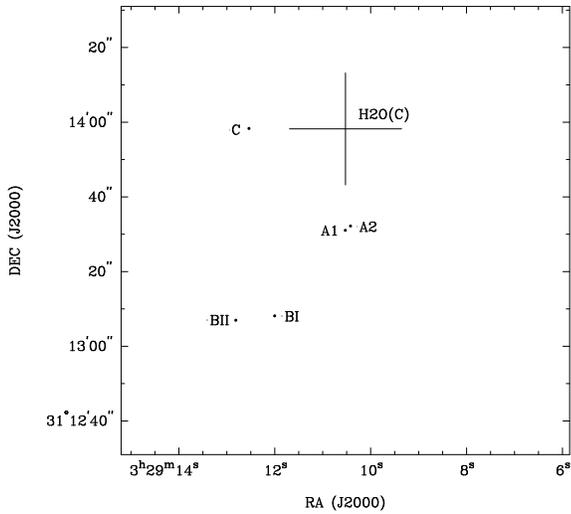}
\caption{
Positions of the known compact sources within the field of view.
{\it Filled circles}:
Radio continuum sources (Looney et al. 2000; Reipurth el al. 2002).
{\it Plus}:
Maser source \HtO(C) (Haschick et al. 1980).
The size of plus corresponds to the uncertainty.}
\end{figure}

\begin{figure}
\includegraphics[width=75mm]{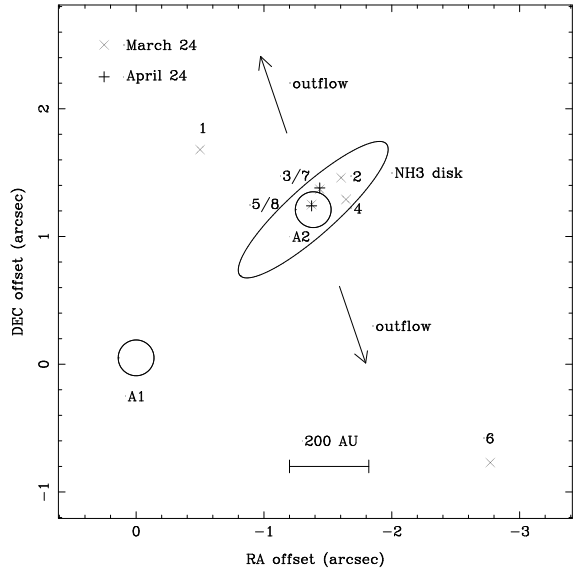}
\caption{
Map of the \HtO\ maser spots in the NGC 1333 IRAS 4A region.
{\it Crosses}:
Maser spots detected on March 24.
PC source numbers are labeled (see Table 1).
The size of markers corresponds to the synthesized beam:
FWHM $\approx$ 0\farcs08.
{\it Pluses}:
Maser spots detected on April 24.
{\it Open circles}:
Positions of the 3.6 cm continuum sources (Reipurth et al. 2002).
The straight line at the bottom
corresponds to 200 AU at a distance of 320 pc.
{\it Arrows}:
Direction of the northeast-southwestern bipolar outflow of IRAS 4A2
(Choi 2005).
{\it Open ellipse}:
Schematic diagram of the NH$_3$ disk of IRAS 4A2.
(Choi et al. 2007).}
\end{figure}

\begin{figure}
\includegraphics[width=75mm]{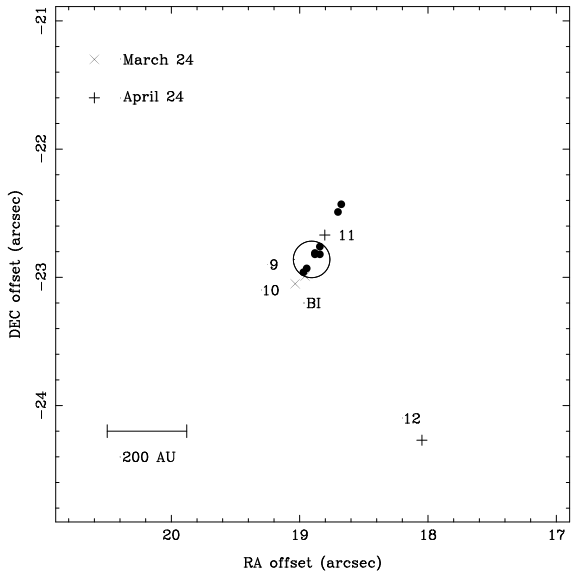}
\caption{
Map of the \HtO\ maser spots in the NGC 1333 IRAS 4BI region.
{\it Crosses}:
Maser spots detected on March 24.
{\it Pluses}:
Maser spots detected on April 24.
{\it Filled circles}:
Maser spots detected by Rodr{\'\i}guez et al. (2002).}
\end{figure}

\begin{figure}
\includegraphics[width=75mm]{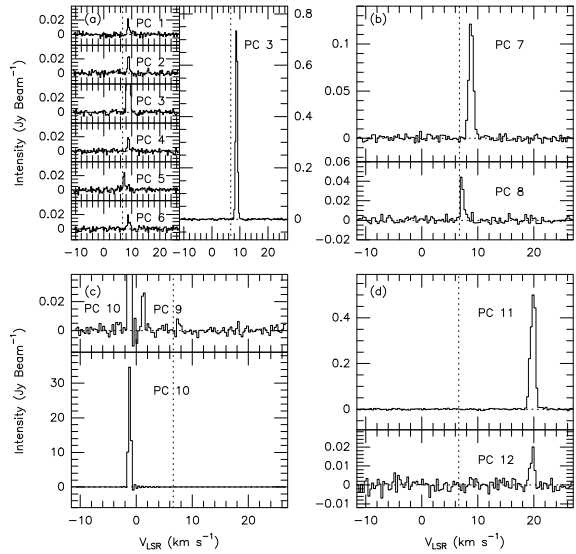}
\caption{
Spectra of the \HtO\ masers.
($a$)
Masers around IRAS 4A2 from the March 2006 observations.
($b$)
Masers around IRAS 4A2 from the April 2006 observations.
($c$)
Masers around IRAS 4BI from the March 2006 observations.
($d$)
Masers around IRAS 4BI from the April 2006 observations.
{\it Vertical dotted line}:
Systemic velocity of the cloud core:
$V_{\rm LSR}$ = 6.7 \kms\ for A2 and 6.6 \kms\ for BI
(Blake et al. 1995; Choi 2001).}
\end{figure}

\begin{figure}
\includegraphics[width=75mm]{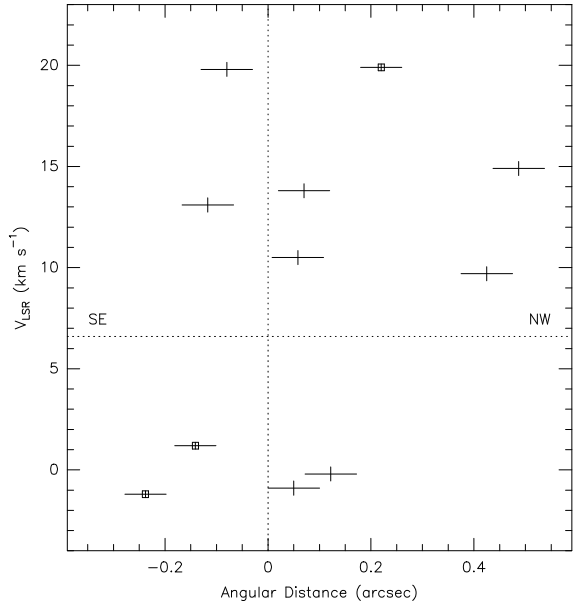}
\caption{
Relation between the maser velocities and the angular distances.
The horizontal axis is
the angular separation between the maser spots and BI
(position of the 3.6 cm source detected by Reipurth el al. 2002),
projected on a best-fit straight line.
{\it Pluses}:
Masers from this work (with squares)
and from Rodr{\'\i}guez et al. (2002) (without squares).
The size of the pluses corresponds to the uncertainties.
{\it Horizontal dotted line}:
Systemic velocity of the cloud core.}
\end{figure}

\end{document}